# INTERPRETING MISSING DATA PATTERNS IN THE ICU
Abstract

Rob O'Shea
robert.o'shea.17@ucl.ac.uk

# Contents

# Abstract

PURPOSE: Clinical examinations are performed on the basis of necessity. However, our decisions to investigate and document are influenced by various other factors, such as workload and preconceptions. Data missingness patterns may contain insights into conscious and unconscious norms of clinical practice. METHODS: We examine data from the SPOTLIGHT study, a multi-centre cohort study of the effect of prompt ICU admission on mortality. We identify missing values and generate an auxiliary dataset indicating the missing entries. We deploy sparse Gaussian Graphical modelling techniques to identify conditional dependencies between the observed data and missingness patterns. We quantify these associations with sparse partial correlation, correcting for multiple collinearity. RESULTS: We identify 35 variables which significantly influence data missingness patterns (alpha = 0.01). We identify reduced recording of essential monitoring such as temperature (partial corr. = -0.0542, p = 6.65e-10), respiratory rate (partial corr. = -0.0437, p = 5.15e-07) and urea (partial corr. = -0.0263, p = 0.001611) in patients with reduced consciousness. We demonstrate reduction of temperature (partial corr. = -0.04, p = 8.5e-06), urine output (partial corr. = -0.05, p = 7.5e-09), lactate (partial corr. = -0.03, p = 0.00032) and bilirubin (partial corr. = -0.03, p = 0.00137) monitoring due to winter pressures. We provide statistical evidence of *Missing Not at Random* patterns in FiO2 and SF ratio recording. CONCLUSIONS: Graphical missingness analysis offers valuable insights into critical care delivery, identifying specific areas for quality improvement.

# Introduction

Missing data is a ubiquitous phenomenon in medical research. Where large numbers of clinical variables are considered for interpretation, the prospect of data absence approaches certainty. Absent data entries are an inevitable source of bias in statistical analysis, as assumptions must ultimately be made in regard to their distribution. Importantly, the generative mechanism for data absence must be considered. If the data is missing completely at random, the missing data distribution may reasonably be assumed to follow the distribution of the observed data entries. In contrast, systematic sources of data missingness may cause imbalanced data recording patterns, biasing the observed data distribution.

Many clinical variables are measured on the basis of clinical necessity. Invasive tests may be omitted at the discretion of the managing clinician, if the pre-test probability of abnormality is deemed negligible. Accordingly, missing entries in these variables typically represent forgone investigations. If such cases, the distribution of the unobserved values is expected to concentrate around the normal range. Conversely, observed values indicate test execution and, consequently, non-negligible pre-test probability of abnormality. Therefore, an amount of data absence in observational research results directly from the clinician's evaluation of the patient. This mechanism of data missingness is influenced directly by the clinician's expectation of the variable's value and is therefore highly likely to follow a *Missing-Not-at-Random* distribution. Management of this missingness pattern is notoriously difficult, as few assumptions may be reasonably made about the distribution of the absent data.

If missingness results from clinical factors, its pattern may be inherently informative. Conversely, if data absence has influenced outcomes, it will function as a confounder if unobserved. In both cases, valuable insights will be gained through examination of the data absence pattern. In fact, several studies have recently demonstrated that data absence patterns may be explicitly modelled to predict outcomes such as mortality (1–3).

We have identified that numerous factors may result in data absence. The extent to which these factors interact with one another is uncertain. Likewise, data absence may conceivably affect many of the variables under investigation, through direct or indirect mechanisms. Furthermore, an uncertain quantity of the absent entries may be generated by actual random mechanisms, resulting in "noise" in the missingness pattern. Each of these considerations complicate the task of data absence modelling.

Graphical modelling methods have elucidated complex systems in fields such as genomics (4,5), neuroscience (6) and engineering (7). Sparse graphical modelling methods such as the graphical lasso (8,9) aim to represent the function of large interactive systems in the simplest possible manner, offering insights into the dependencies of individual variables. Graphical modelling methods have recently been deployed for the purpose of missingness analysis, allowing the identification of variables which influence, or are influenced by, data recording practices (10,11). We apply graphical data missingness modelling to a dataset of intensive care observations. We explicitly model missingness patterns, by generating binary indicators of data absence for each variable and including them as additional variables. We learn an undirected Gaussian Graphical Model from the data using the graphical lasso (8,9). We identify the conditional dependencies of these variables and extract effect sizes from the partial correlation matrix.

# Methodology

## Data

We examine an observational dataset, containing physiological, biochemical and administrative data on 12,495 ICU admissions in 48 UK centres. This data was collected during the (SPOT)light (Sepsis Pathophysiological and Organisational Timing) study (12), an prospective study of the mortality reduction associated with early intervention in septic patients in the intensive care unit. All variables except mortality outcomes were collected simultaneously during ICU admission. Mortality outcomes were collected retrospectively. As part of the original study protocol, patients missing large amounts of routine data, such as those who died during evaluation, were excluded from the study. Patients younger than 18 years of age and elective admissions were also excluded.

## Modelling Approach

For each variable with missing values, a binary indicator variable was generated as follows:

$$f(x_i) = \begin{cases} 1, & i^{th} \text{ entry is present in } x \\ 0, & i^{th} \text{ entry is missing in } x \end{cases} \quad (1)$$

Hereafter, we refer to the original variables as *observation variables* and the binary indicators of data absence as *completeness variables*. The dataset of observation variables was concatenated with the completeness variables. Subsequently, all missing values were randomly imputed with samples of the observed entries in the same variables. Therefore, the imputed values were expected to have independent distribution (This statement is proved in lines 4-9 of the theoretical analysis). 25 imputed datasets were generated with this method.

| Variable | Category | Missing Proportion |
|---|---|---|
| HR | Vital Physiology | 0.016 |
| SBP | Vital Physiology | 0.024 |
| DBP | Vital Physiology | 0.035 |
| Temp | Vital Physiology | 0.095 |
| RR | Vital Physiology | 0.027 |
| FiO2 | Vital Physiology | 0.036 |
| PF ratio | Vital Physiology | 0.624 |
| uvol/1h | Vital Physiology | 0.428 |
| SF ratio | Vital Physiology | 0.057 |
| AVPU | Vital Physiology | 0.093 |
| pH | Blood Tests | 0.59 |
| Sodium | Blood Tests | 0.137 |
| WCC | Blood Tests | 0.149 |
| Urea | Blood Tests | 0.159 |
| Creatinine | Blood Tests | 0.137 |
| Platelets | Blood Tests | 0.154 |
| Bilirubin | Blood Tests | 0.39 |
| Lactate | Blood Tests | 0.727 |
| Age | Demographics | 0 |
| Male | Demographics | 0 |
| Died (7 days) | Mortality | 0 |

| Died (28 days) | Mortality | 0 |
| Died (90 days) | Mortality | 0 |

*Table 1: Variable categories and completeness*

All variables were transformed to a multivariate sub-Gaussian distribution with Liu's non-paranormal transformation (13), implemented by the R package *huge* (14). Undirected Gaussian graphical models were on each imputed dataset using Jankova's de-sparsified graphical lasso (15), implemented with the R package *SILGGM* (16). The graphical lasso (8,9) solves for a sparse inverse covariance matrix $\mathbf{\Sigma}^{-1}$ by solving the following optimisation problem:

$$\widehat{\mathbf{\Sigma}^{-1}} := \arg\min \left\{ trace\left(\mathbf{\Sigma}^{-1^\mathrm{T}} \widehat{\mathbf{\Sigma}}\right) - \log \det(\mathbf{\Sigma}^{-1}) + \lambda \|\mathbf{\Sigma}^{-1}\|_{1,OFF} \right\} \qquad (2)$$

Where $\widehat{\mathbf{\Sigma}}$ is the sample covariance matrix, $\lambda$ is a regularisation parameter and $\|\cdot\|_{1,OFF}$ is the off-diagonal $l_1$ norm. We determine the optimal $\lambda$ for each dataset according to Lysen's Rotation Invariance Criterion (13), implemented with *huge*. Partial correlation matrices were extracted from each model. The partial correlation matrix is computed from the inverse covariance matrix as follows:

$$\frac{\mathbf{\Sigma}^{-1}_{ij}}{\mathbf{\Sigma}^{-1}_{ii} \mathbf{\Sigma}^{-1}_{jj}} = \rho_{X_i, X_j | X_{\setminus \{i,j\}}}$$

Graphical modelling methods exploit the variable selection properties of sparse matrix inversion, as variables $X_i$ and $X_j$ are found to be conditionally dependent if and only if $\mathbf{\Sigma}^{-1}_{ij} \neq 0$. Therefore, conditional dependencies between completeness variables and observation variables may be identified by non-zero entries in the inverse covariance matrix (19). Partial correlations from each model estimate were transformed to Z-scores before mean pooling, then re-transformed to correlation coefficients. This operation was performed as follows (note: here Σ denotes the summation function rather than the covariance matrix):

$$\rho_{pooled} = \tanh\left(\frac{\sum_{i=1}^{n_{imputes}} \mathrm{atanh}(\rho_i)}{n_{imputes}}\right) \qquad (3)$$

Correlation significance was subsequently computed by fisher transformation of the correlation coefficients. Significant arcs were identified with an alpha threshold of 0.01. All arcs connecting observation variables to missingness variables were extracted. For each arc connecting an observation variable with a missingness variable, we also extracted the partial correlation of the corresponding pair of observation variables.

# Results

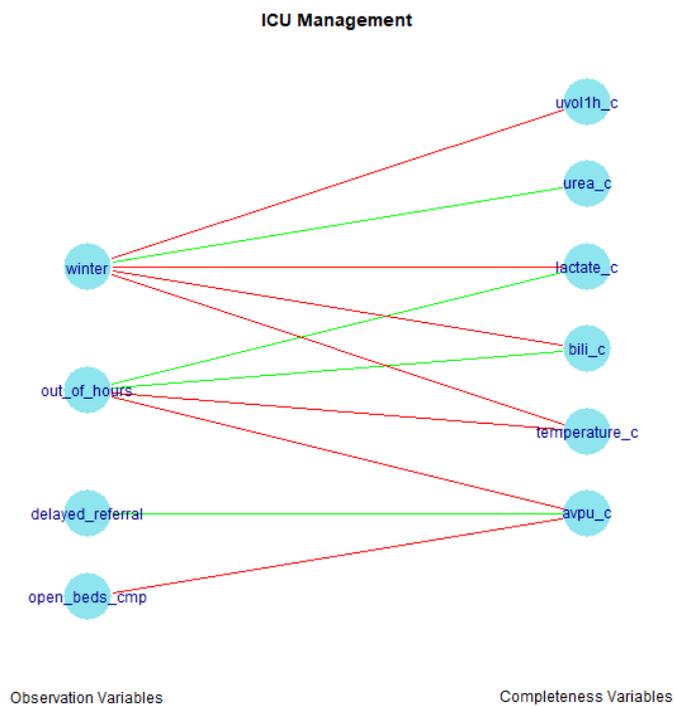

*Figure 1: Conditional dependencies between observation variables in the ICU management category (left column) and completeness variables (right column). Green arcs indicate a positive associations – such that increasing values of the observation variable are associated with increased completeness of the completeness variable. Red arcs indicate negative associations.*

## ICU Management

Lower completeness of temperature (partial corr. = -0.04, p = 8.5e-06), urine output (partial corr. = -0.05, p = 7.5e-09), lactate (partial corr. = -0.03, p = 0.00032) and bilirubin (partial corr. = -0.03, p = 0.00137) was attained during winter. Serum urea completeness increased during this period (partial corr. = 0.05, p = 2.1e-07). Out-of-hours admissions had lower temperature completeness (partial corr. = -0.04, p = 1.6e-05) and AVPU score completeness (partial corr. = -0.05, p = 8.9e-08). Lactate completeness increased in out-of-hours admissions (partial corr. = 0.03, p = 0.00117), as did bilirubin completeness (partial corr. = 0.03, p = 3.21e-04). Recording of AVPU completeness was higher in referrals which were classified as delayed by the clerking doctor (partial corr. = 0.03, p = 0.00116), though recorded scores were equivalent (partial corr. = 0, p = 0.46117). Unexpectedly, AVPU score completeness decreased (partial corr. = -0.08, p = < 2e-16) during periods of high bed availability. A positive partial correlation was found between recorded values and the number of open beds (partial corr. = 0.02, p = 0.00685), indicating that scores were higher, on average, during times of high bed availability.

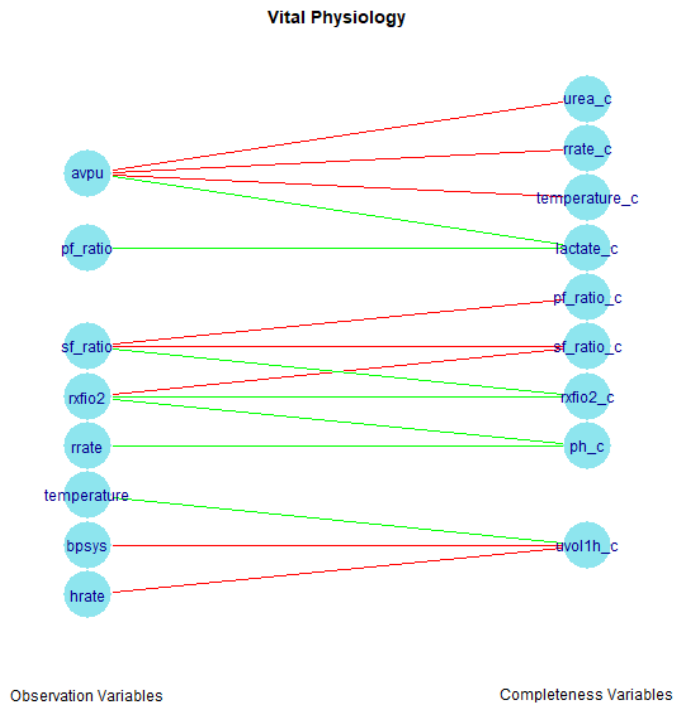

*Figure 2: Conditional dependencies between observation variables in the vital physiology category (left column) and completeness variables (right column). Green arcs indicate a positive associations – such that increasing values of the observation variable are associated with increased completeness of the completeness variable. Red arcs indicate negative associations.*

## Vital Physiology

Urine output completeness correlated inversely with systolic blood pressure (partial corr. = -0.0382, p = 9.57e-06), inversely with heart rate (partial corr. = -0.0249, p = 0.002720) and positively with body temperature (partial corr. = 0.0236, p = 0.004147). Recorded urine output values correlated positively with systolic blood pressure (partial corr. = 0.0539, p = 7.89e-10) and body temperature (partial corr. = 0.0404, p = 3.15e-06). pH documentation correlated with respiratory rate (partial corr. = 0.0389, p = 6.75e-06) . Recorded pH values also correlated positively with respiratory rate (partial corr. = 0.0208, p = 0.009999). Interestingly, a positive partial correlation was found between FiO2 completeness and FiO2 levels (partial corr. = 0.0924, p = < 2e-16). pH correlated positively with FiO2 completeness (partial corr. = 0.0395, p = 5.08e-06) and inversely with FiO2 level (partial corr. = 0.0208, p = 0.009999).

SF ratio completeness correlated inversely with FiO2 (partial corr. = -0.1483, p = < 2e-16) as did decorded SF ratio values (partial corr. = -0.7086, p = < 2e-16). Interestingly, SF ratio completeness was found to correlate inversely with SF ratio values (partial corr. = -0.0399, p = 4.13e-06).  SF ratio correlated positively with FiO2 completeness (partial corr. = 0.068, p = 1.33e-14), and inversely with PF ratio completeness (partial corr. = -0.0995, p = < 2e-16). Lactate completeness correlated with PF ratio (partial corr. = 0.0463, p = 1.08e-07). Patients with decreased conscious levels were less likely to have temperature (partial corr. = -0.0542, p = 6.65e-10), respiratory rate (partial corr. = -0.0437, p = 5.15e-07) and  urea (partial corr. = -0.0263, p = 0.001611) documented. AVPU scores correlated inversely with recorded respiratory rate (partial corr. = -0.0437, p = 5.15e-07) and temperature (partial corr. = -0.0248, p = 0.002750) values. Patients with decreased consciousness were more likely to have lactate documented (partial corr. = 0.0246, p = 0.002959).

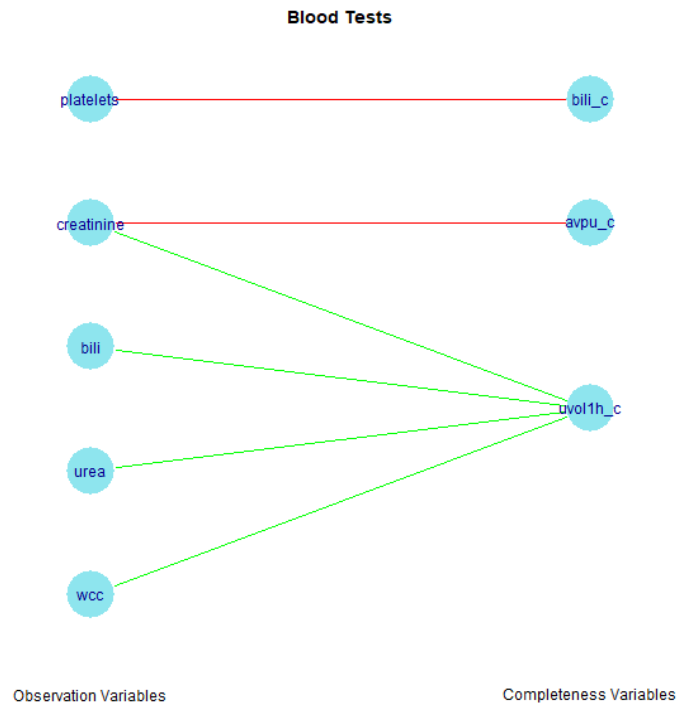

*Figure 3: Conditional dependencies between observation variables in the blood test category (left column) and completeness variables (right column). Green arcs indicate a positive associations – such that increasing values of the observation variable are associated with increased completeness of the completeness variable. Red arcs indicate negative associations.*

## Blood Tests

Urine output data recording was greater in patients with high leukocyte counts (partial corr. = 0.0361, p = 2.64e-05), urea (partial corr. = 0.0455, p = 1.81e-07) and creatinine (partial corr. = 0.0381, p = 1.01e-05). Recorded urine output values inversely correlated with creatinine level (partial corr. = -0.0618, p = 2.31e-12). AVPU score completeness decreased in patients with raised creatinine levels (partial corr. = -0.034, p = 7.31e-05). A negative partial correlation was found between platelet count completeness and serum bilirubin levels (partial corr. = -0.0237, p = 0.004060). Recorded platelet count values demonstrated a strong negative partial correlation with recorded bilirubin levels (partial corr. = -0.1935, p = < 2e-16). Urine output documentation was increased in bilirubinaemic patients (partial corr. = 0.0421, p = 1.24e-06).

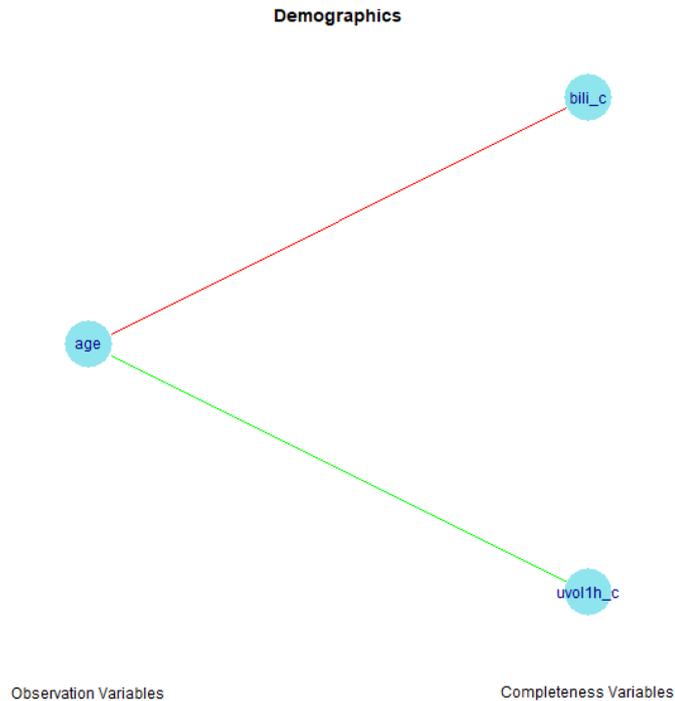

*Figure 4: Conditional dependencies between observation variables in the demographics category (left column) and completeness variables (right column). Green arcs indicate a positive associations – such that increasing values of the observation variable are associated with increased completeness of the completeness variable. Red arcs indicate negative associations.*

### Demographics

Older patients were more likely to have urine output recorded (partial corr. = 0.0362, p = 2.61e-05) and less likely to have serum bilirubin recorded (partial corr. = -0.075, p = < 2e-16).

## Discussion

### Causal Assumptions

35 associations were identified between observation variables and completeness variables. In order to deduce causality from such associations, several assumptions must be made (20,21). Firstly, the sample data must be representative of the true underlying distribution. This study investigates a highly specific population, adult ICU patients. This population has the atypical feature that all individuals have some critical illness. This restriction results in unusual phenomena, such as an apparent negative correlation between body temperature and 7-day mortality (partial corr. = -0.037, p = 1.57e-05) indicating a protective effect of fever. This association presumably results from increased mortality observed in non-infectious critical illness. Secondly, it is assumed that all relevant variables either observed or distributed randomly with respect to the observation variables. Exhaustive measurement of all relevant variables is not feasible in medical research, and causal deductions therefore assume approximately random distribution of unobserved variables.

Given the temporal sequence of data collection and observed events, causal direction could be reasonably assumed for arcs between observation variables and completeness variables. All observation variables except mortality outcomes were fixed prior to data recording. Therefore, completeness variables were exclusively effects in associations with observation variables. Mortality was documented retrospectively and would be the effect in any association with non-mortality completeness variables. However, no such association was found in this study.

### Specific Associations

Reduced data completeness was observed across a wide range of variables during wintertime. This broad effect probably resulted from increased clinical workload, indicated by the decrease in bed availability during wintertime (partial corr. = -0.21, p = < 2e-16). Completeness of standard septic monitoring such as temperature, lactate and urine output decreased during the winter season. These deficits warrant further monitoring, in order to ensure consistent delivery of sepsis care during winter pressures. Essential monitoring such as temperature and AVPU score decreased out-of-hours, indicating decreased thoroughness of clinical monitoring. The decrease in AVPU documentation during periods of high bed availability paired with the increase in the recorded scores during this time indicates that recording of normal AVPU scores decreased during periods of high bed availability. This may be due to some relaxation of vital monitoring during quiet periods. As AVPU scores demonstrated strong positive partial correlation with 7-day mortality (partial corr. = 0.033, p = 1.22e-04), such laxity may compromise crucial clinical care. Completeness of temperature, respiratory rate and urea variables were reduced in patients with decreased conscious levels. As recorded temperatures correlated with conscious level and recorded respiratory rates correlated inversely with conscious level, it is likely that these measurements were forgone due to were expected to anticipated normality. However, vital monitoring is a fundamental component of critical care, and this deficit warrants further investigation.

FiO2 and SF ratio completeness correlated with FiO2 values and SF ratio values respectively. This interesting phenomenon demonstrates the value of graphical missingness analysis. As missing entries in any observation variable were filled with random samples of the observed entries, no observation variable was expected to correlate with its corresponding missingness variable. However, in the case that the observation variable and its missingness variable both correlate strongly with a third variable (or multiple other variables) a partial correlation may be found between them. This provides strong evidence that the data absence risk depended on the missing value (this result is proved in lines 11-15 of the theoretical analysis). High FiO2 levels were more likely to be documented, as were low SF ratios. These associations reflect normal clinical documentation practice, where normal observations are sometimes omitted for brevity.

A strong negative partial correlation was identified between recorded platelet counts and serum bilirubin levels (partial corr. = -0.0237, p = 0.004060). This association may be due to increased platelet aggregation observed in hyperbilirubinaemia (22,23). However, it is important to note that completeness of platelet counts decreased in hyperbilirubinaemic patients (partial corr. = -0.0237, p = 0.004060). Therefore, patients at increased risk of thrombocytopaenia were less likely to have platelet count recorded. A significant inverse partial correlation was identified between platelet count and 7-day mortality in this data (partial corr. = -0.0258, p = 1.99e-03). Meticulous documentation of platelet counts in hyperbilirubinaemic patients will likely lead to increased detection of thrombocytopaenia, an independent predictor of ICU mortality in this study.

### Strengths and Weaknesses

This data was collected for the purpose of assessing the effect of prompt admission on mortality in septic ICU patients. Therefore, clinicians were unaware of any prospect of data missingness pattern analysis. This conferred perfect blinding during data collection, strengthening the objectivity of the results. The study recruited a large population in multiple centres. Therefore, the findings of this analysis are likely generalisable to all UK ICU units.

Several important variables, including comorbidities, medications and genetic factors may have confounded the results of this study. For the purposes of this analysis, these factors were assumed to be approximately randomly distributed, and that their influence on study results was negligible. The positive partial correlation found between serum bilirubin completeness and the out-of-hours variable is not easily explained. This result may be due latent confounding, such as an increase in the number of patients with abdominal complaints in the evenings. Alternatively, this may be an example of a false positive result, due to under regularisation during graph inference. However, the majority of associations found by the analysis were consistent with known clinical phenomena.

## Conclusions

Graphical modelling provided several insights into data missingness patterns in this study. In many cases the missingness associations concur with norms of clinical practice, where medical investigation is performed on the basis of necessity. Additionally, several missingness associations identified areas for quality improvement in clinical monitoring and documentation. Graphical modelling allowed the deduction of several Missing-Not-at-Random patterns, a previously intractable task. This approach is general and further insights may be gained by its application to other clinical datasets.

## Supplementary Material - Theoretical Analysis

Let $X$ be a $p$-dimensional multivariate Gaussian vector, with zero mean and unit variance such that $X \sim N_p(0,1)$. Let there be $n$ observations of $X$. Let $a$ be a partially observed variable in $X$. Let $c$ be a binary variable indicating the observation of $a$, such that:

$$c_i = \begin{cases} 1, & a_i \text{ is observed} \\ 0, & a_i \text{ is missing} \end{cases} \quad (1)$$

Let $Z$ be the set of remaining variables in $X$. Let $M$ denote the missing indices in $a$, such that:

$$M = \{i : i \in \{1, \ldots, n\}, c_i = 1\} \quad (2)$$

Let the missing entries of $a$ be imputed with a random selection of the observed entries of $a$, such that:

$$a_i \leftarrow a_j, i \in M, j \notin M \quad (3)$$

$$\rho_{a,c} = \frac{\langle a, c \rangle}{\langle a, a \rangle \langle c, c \rangle} = \frac{\langle a_M, c_M \rangle + \langle a_{\setminus M}, c_{\setminus M} \rangle}{\langle a, a \rangle \langle c, c \rangle} \quad (4)$$

$$c_i = \begin{cases} 1, & i \in M \\ 0, & i \notin M \end{cases} \quad (5)$$

$$\therefore \rho_{a,c} = \frac{\langle a_M, 1 \rangle + \langle a_{\setminus M}, 0 \rangle}{\langle a, a \rangle \langle c, c \rangle} \quad (6)$$

As $a_M \sim N(0,1)$:

$$\langle a_M, 1 \rangle = 0 \quad (7)$$

$$\therefore \rho_{a,c} = \frac{0 + 0}{\langle a, a \rangle \langle c, c \rangle} = 0 \quad (8)$$

Consider a conditional dependency between $c$ and $Z$, given $a$.

$$\rho_{c,Z|a} = \frac{\rho_{c,Z} - \rho_{a,Z} \rho_{a,c}}{\sqrt{(1 - \rho^2_{a,Z})} \sqrt{(1 - \rho^2_{a,c})}} \quad (9)$$

$$\therefore \rho_{c,Z|a} = \frac{\rho_{Z,c}}{\sqrt{(1 - \rho^2_{a,Z})}} \quad (10)$$

Consider the case of a non-zero conditional dependency between $a$ and $c$, given $Z$.

$$\rho_{a,c|Z} \neq 0 \quad (11)$$

$$\therefore \frac{\rho_{a,c} - \rho_{a,Z} \rho_{c,Z}}{\sqrt{(1 - \rho^2_{a,Z})} \sqrt{(1 - \rho^2_{c,Z})}} \neq 0 \quad (12)$$

$$\therefore \frac{0 - \rho_{a,Z} \rho_{c,Z}}{\sqrt{(1 - \rho^2_{a,Z})} \sqrt{(1 - \rho^2_{c,Z})}} \neq 0 \quad (13)$$

$$\therefore \rho_{a,Z}\rho_{c,Z} \neq 0 \tag{14}$$

$$\therefore \rho_{a,Z} \neq 0, \rho_{c,Z} \neq 0 \tag{15}$$

# References

1.  Sesen MB, Nicholson AE, Banares-Alcantara R, Kadir T, Brady M. Bayesian Networks for Clinical Decision Support in Lung Cancer Care. PLoS One. 2013;8(12):82349.

2.  Nagrebetsky A, Bittner EA. Missing Data and ICU Mortality Prediction. Crit Care Med. 2017 Dec;45(12):2108–9.

3.  Sharafoddini A, Dubin JA, Maslove DM, Lee J. A New Insight Into Missing Data in Intensive Care Unit Patient Profiles: Observational Study. JMIR Med Informatics. 2019 Jan 8;7(1):e11605.

4.  Hawe JS, Theis FJ, Heinig M. Inferring Interaction Networks From Multi-Omics Data. Front Genet. 2019 Jun 12;10:535.

5.  Liu Q, Ihler A. Learning scale free networks by reweighted L1 regularization. In: Journal of Machine Learning Research. 2011. p. 40–8.

6.  Bielczyk NZ, Uithol S, van Mourik T, Anderson P, Glennon JC, Buitelaar JK. Disentangling causal webs in the brain using functional magnetic resonance imaging: A review of current approaches. Netw Neurosci (Cambridge, Mass). 2019;3(2):237–73.

7.  Tang K, Parsons DJ, Jude S. Comparison of automatic and guided learning for Bayesian networks to analyse pipe failures in the water distribution system. Reliab Eng Syst Saf. 2019 Jun 1;186:24–36.

8.  Friedman J, Hastie T, Tibshirani R. Sparse inverse covariance estimation with the graphical lasso. 2007;1–14.

9.  Yuan M. High Dimensional Inverse Covariance Matrix Estimation via Linear Programming. Vol. 11, Journal of Machine Learning Research. 2010.

10. Mohan K, Pearl J, Tian J. Graphical models for inference with missing data. In: Advances in Neural Information Processing Systems. 2013. p. 1277–85.

11. Kolar M, Xing EP. Estimating sparse precision matrices from data with missing values. In: Proceedings of the 29th International Conference on Machine Learning, ICML 2012. 2012. p. 551–8.

12. Harris S, Singer M, Sanderson C, Grieve R, Harrison D, Rowan K. Impact on mortality of prompt admission to critical care for deteriorating ward patients: an instrumental variable analysis using critical care bed strain. Intensive Care Med. 2018 May 7;44(5):606–15.

13. Liu H, Han F, Yuan M, Lafferty J, Wasserman L. The Nonparanormal skeptic. 2012.

14. Zhao, Liu L, Roeder K. The huge Package for High-dimensional Undirected Graph Estimation in R. J Mach Learn Res 2012 April ; 13 1059–1062. 2012;April(13):1059–62.

15. Janková J, van de Geer S. Confidence intervals for high-dimensional inverse covariance estimation. Electron J Stat. 2015;9:1205–29.

16. Zhang R, Ren Z, Chen W. SILGGM: An extensive R package for efficient statistical inference in large-scale gene networks. PLoS Comput Biol. 2018;14(8).

17. Lysen S. Permuted Inclusion Criterion: A Variable Selection Technique. Publicly Access Penn Diss. 2009;

18. Zhao T, Roeder K, Lafferty J, Wasserman L. The huge Package for High-dimensional Undirected Graph Estimation in R. 2012;1–12.

19. Lauritzen S. Elements of Graphical Models. 2011;(1):1–13.

20. Pearl J. An Introduction to Causal Inference. Int J Biostat. 2010;6(2):7.

21. Lauritzen SL. Graphical Models. Oxford University Press; 1996.

22. Kundur AR, Singh I, Bulmer AC. Bilirubin, platelet activation and heart disease: A missing link to cardiovascular protection in Gilbert's syndrome? Atherosclerosis. 2015 Mar 1;239(1):73–84.

23. Pasko V, Batushkin V. Platelet aggregation in patients with heart failure of ischemic origin depends on blood levels of iron and bilirubin. Atherosclerosis. 2018 Aug 1;275:e145–6.